\documentclass
[twocolumn,showpacs,preprintnumbers,amsmath,amssymb]{revtex4}
\usepackage{graphicx}
\usepackage{dcolumn} 
\usepackage{bm}    
\begin{document}
\title{Transverse mode coupling instability of the bunch\\
with oscillating wake field and space charge}
\author{V. Balbekov}
\affiliation {Fermi National Accelerator Laboratory\\
P.O. Box 500, Batavia, Illinois 60510}
\email{balbekov@fnal.gov} 
\date{\today}

\begin{abstract}

Transverse mode coupling instability of a single bunch
caused by oscillating wake field is considered in the paper.
The instability threshold is found at different frequencies of the wake
with space charge tune shift taken into account.
The wake phase advance in the bunch length from 0 up to $4\pi$ 
is investigated.  
It is shown that the space charge can push the instability threshold
up or down dependent on the phase advance. 
Transition region is investigated thoroughly,
and simple asymptotic formulas for the threshold are represented.

\end{abstract}
\maketitle

\section{INTRODUCTION}

The transverse mode coupling instability (TMCI) of a bunch with both wake 
field (WF) and space charge (SC) was first considered in paper \cite{Bl1}. 
It has been shown there that, at a moderate ratio of the SC tune shift to the 
synchrotron tune $(\Delta Q/Q_s)$, the SC pushes up threshold of the instability 
caused by a negative wake.
Hereafter the result has been confirmed in papers \cite{Bu1}-\cite{Bl2}.
 
However, more confusing picture appears at the larger value of this ratio 
like a hundred or over.
It has been suggested in Ref.~\cite{Bu1} that the threshold growth ceases 
behind this border coming to 0 at $\,\Delta Q/Q_s\rightarrow\infty$.  
By contrast, it was asserted in Ref.~\cite{Ba1} that negative wake cannot 
excite the TMCI in this limiting case. 

An explanation of the collision has been proposed in Ref.~\cite{Ba2} where
it has been noted that approximate methods of solution were applied in all 
mentioned articles.
Typically, an expansion of the bunch coherent displacement in terms of some series
of basic functions was used there, with subsequent truncation of the series. 
It turns out that, with this approach, the TMCI threshold can strongly and not 
monotonously depend from the actual number of used basic functions.

The problem has been clarified in recent publication \cite{Ba3} where 
the method of solution has been developed which does not use the expansion 
technique at all, and therefore is applicable at arbitrary SC tune shift.
In particular, it has been shown that the TMCI threshold of negative wakes 
is asymptotically proportional to $\,\Delta Q/Q_s$ in contrast with the case 
of positive wake whose threshold is $\,\propto Q_s/\Delta Q$.
The statement has been extended in the paper on any monotonous WF, e.g. on the 
resistive wall impedance.

In the presented paper, this method is applied for investigation of
oscillating (resonant) wake fields.
It is shown that the SC tune shift can suppress or intensify the instability, 
dependent on the wake phase advance from the bunch head to its tail. 
It is shown as well that, at some conditions, the dependence of the TMCI rate 
on the wake amplitude is a non-monotonic function.

Some of these problems were considered earlier using other models and techniques 
of the calculations  \cite{Ba2}, \cite {Bu2}. 
All the results are in rather good consent to be convinced of applicability
of the used methods.
 Their detailed comparison is carried out at the end of Sec. V.
\section{PHYSICAL MODEL}

Following \cite{Ba3}, we represent the bunch transverse coherent 
displacement in the rest frame  as  the part of the function
\begin{eqnarray}
\bar X(z,t) = \bar Y(z)\exp\big[-iQ_\beta(z/R+\Omega_0t)-i\Omega_0\nu t\big] 
\end{eqnarray}
where $\,z$ is the longitudinal coordinate, $\,t\,$ is time, $\,R$ is the 
machine radius, $\,\Omega_0$ is the revolution frequency, 
$\,Q_\beta $ and $\,\nu$ are the central betatron tune and the addition to 
it due to the wake field. 
The machine chromaticity in not included in the expression as a factor of 
a small importance for the TMCI threshold \cite{Ba2}.

In framework of the model, it is convenient to use another longitudinal 
coordinate $\,\vartheta\propto z$ which is adjusted to the bunch located
on the interval $\,0\le\vartheta\le\pi$.  
Then the rearranged function $\,\bar Y$ satisfies the equation 
\begin{eqnarray}
 Q_{s0}^2\bar Y''(\vartheta)+\nu\hat\nu\bar Y(\vartheta)=
\frac{2\hat\nu}{\pi}\int_\vartheta^\pi q(\vartheta'-\vartheta)\bar Y(\vartheta')
\,d\vartheta'
\end{eqnarray}
with the boundary conditions 
\begin{eqnarray}
\bar Y'(0)=\bar Y'(\pi) = 0
\end{eqnarray}
where $\,Q_{s0}$ is the synchrotron tune, and  $\,\hat\nu=\nu+\Delta Q$ 
with $\,\Delta Q$ as the space charge tune shift \cite{Ba3}.
The function $\,q(\vartheta)$ is proportional to the usual transverse wake 
function at corresponding distance $\,W_1(z_\vartheta)$:
\begin{eqnarray}
q(\vartheta) = \frac{r_0RN_bW_1}{8\pi\beta^2\gamma Q_\beta}
\end{eqnarray}
where $\,r_0=e^2/mc^2$ is the particle electromagnetic radius, $\,N_b$ 
is the bunch population, $\,\beta$ and $\,\gamma$ are the normalized 
velocity and energy of the particles.

The oscillating wake of the form
\begin{eqnarray}
q(\vartheta) =  q_0\exp(-\alpha\theta)\cos\kappa\vartheta
\end{eqnarray}
will be considered in this paper. 
Then Eq.~(2) obtains the form
\begin{eqnarray}
 \bar Y''(\vartheta)+{\cal P}\bar Y(\vartheta)=\nonumber\\
\frac{2{\cal Q}}{\pi}\exp(\alpha\theta)
\int_\vartheta^\pi \exp(-\alpha\theta')\cos\kappa(\vartheta'-\vartheta)
\bar Y(\theta')\,d\vartheta'
\end{eqnarray}
with the notations
\begin{eqnarray}
 {\cal P}=\frac{\nu\hat\nu}{Q_{s0}^2}, 
 \qquad {\cal Q}=\frac{q_0\hat\nu}{Q_{s0}^2}
\end{eqnarray}
It is easy to see that the integral-differential Eq.~(6) 
is reducible to the net differential equation of $4^{th}$ order:
\begin{eqnarray}
 \bar Y^{IV}-\alpha\bar Y'''+({\cal P}+\kappa^2)\bar Y''+\nonumber\\
\left(\frac{2{\cal Q}}{\pi}-\alpha{\cal P}\right)\bar Y'
+\kappa^2{\cal P}\bar Y=0
\end{eqnarray}
with boundary conditions given by Eq.~(3) and more 
\begin{eqnarray}
 \bar Y''(\pi)=-{\cal P}\bar Y(\pi), \qquad \bar Y'''(\pi)=
-\frac{2{\cal Q}}{\pi}\,\bar Y(\pi).
\end{eqnarray}
Similar in appearance equations of third order were considered 
in Ref.~\cite{Ba3}. 
Like them, it is possible to solve Eq.~(8) step-by-step starting from the 
point $\,\vartheta=\pi$ with initial conditions
\begin{subequations}
\begin{eqnarray}
 \bar Y(\pi)=1, \qquad \bar Y'(\pi)=0, \\
 \bar Y''(\pi)=-{\cal P}, \qquad  \bar Y'''(\pi)=\!-\frac{{\cal Q}}{\pi}
\end{eqnarray}
\end{subequations}
and going down to the point $\,\vartheta=0$. 
Some trial value of $\,\cal{P}$ 
has to be used each time with chosen parameters  $\,\kappa,\;\alpha$ and 
$\,{\cal Q}$; however, only those of them resulting in the relation  
$\,Y'(0)=0$ should be taken as the valid solutions satisfying all the required 
conditions.
With any $\,\cal Q$, there is an infinite discrete sequence of the 
{\it eigennumbers} $\;{\cal P}_n$ satisfying these conditions.

\section{EIGENNUMBERS OF THE EQUATION}

It is the general requirement for applicability of the single bunch 
approximation that the damping coefficient is large enough to neglect 
the bunch-to-bunch or the turn-by-turn interaction. 
However, it does not exclude that the damping is negligible within the bunch: 
$\,\exp(-\pi\alpha)\simeq 1$.
Therefore the case $\,\alpha=0$ is considered for the beginning.
\\

Several eigennumbers  of Eq.~(8) are represented in Figs.~1-3 where different 
frequencies of the wake are considered (the used value of $\,\kappa$ is designated 
right in the graph).
Six lowest real eigennumbers $\,{\cal P}_n$ are plotted against $\,{\cal Q}$ in 
each the case.
As one would expect, $\,{\cal P}_n=n^2$ at $\,{\cal Q}=0$ independently 
on $\,\kappa$.
The index $\,n$ will be used further to mark the lines.
\

The overall behavior of the plots considerably depends on the frequency.
The upper graph $(\kappa=0.5) $ is similar to the case of rectangular 
wake $\,\kappa=0$, both qualitative and quantitative \cite{Ba3}.
The next case $\,\kappa=1$ has a different appearance because its lowest 
line $\,n=0$ outspreads far to the right where it connects with the upper 
line $\,n=5$.
Figure 3 with $\,\kappa=1.5$ differs ever stronger because it contains the 
lines $\,n=0$ and $\,n=1$ starting together in the point $\,{\cal Q}\simeq-1$ 
and separately going to the right.   
 \begin{figure}
 \includegraphics[width=80mm]{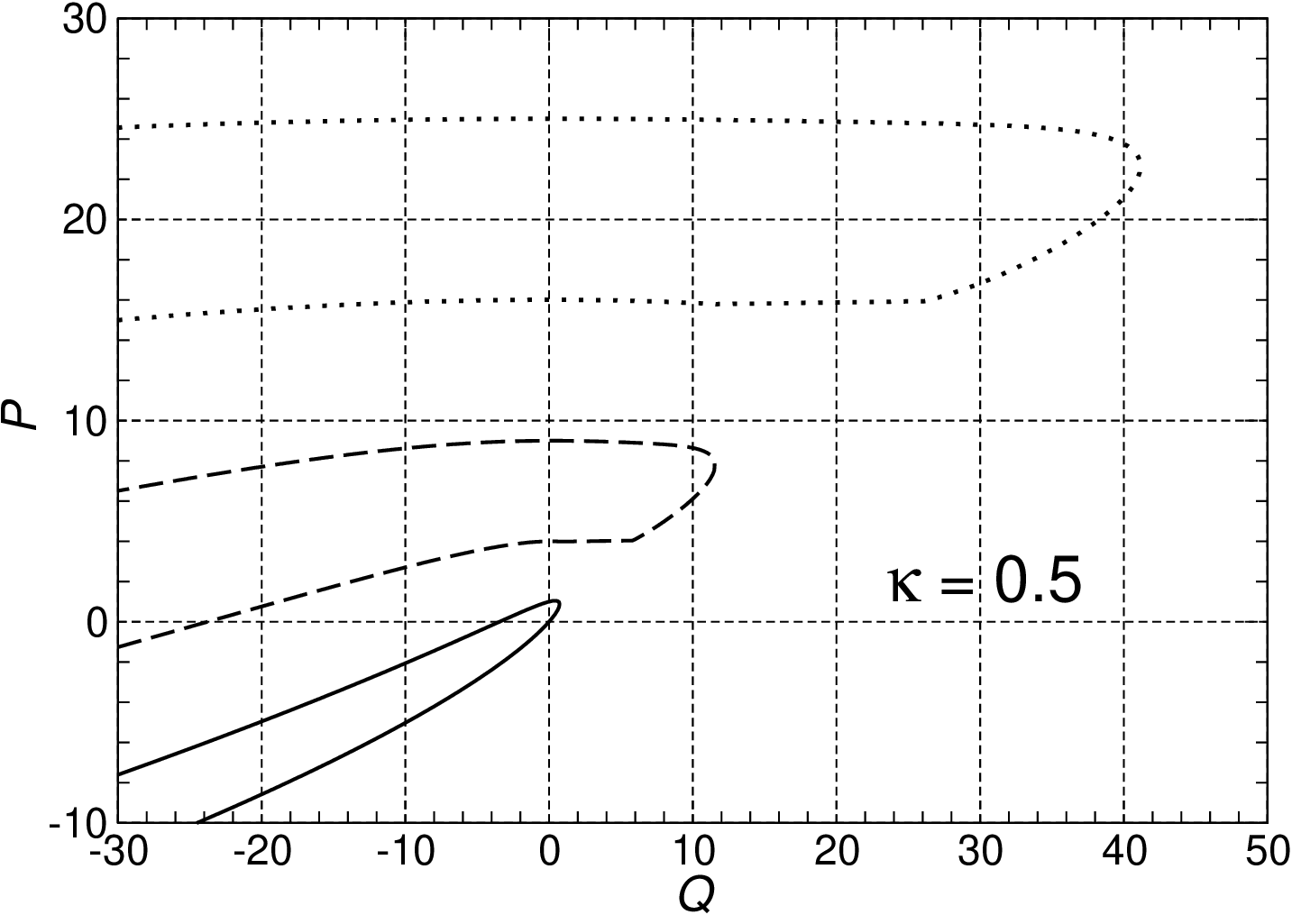}
\caption{Several lowest real eigennumbers of Eq.~(6) and Eq.~(8) at $\,
\alpha=0,\;\kappa=0.5$.
The picture insubstantially differs from the case of rectangular wake 
$\,\kappa=0$ \cite{Ba3}. 
In particular, numbers of the coalescence lines are: 
$\,n=$ (0,1); (2,3); (4,5); etc.
All the lines diverge at $\,{\cal Q}\rightarrow -\infty.$} 
 \end{figure}
\begin{figure}
 \includegraphics[width=80mm]{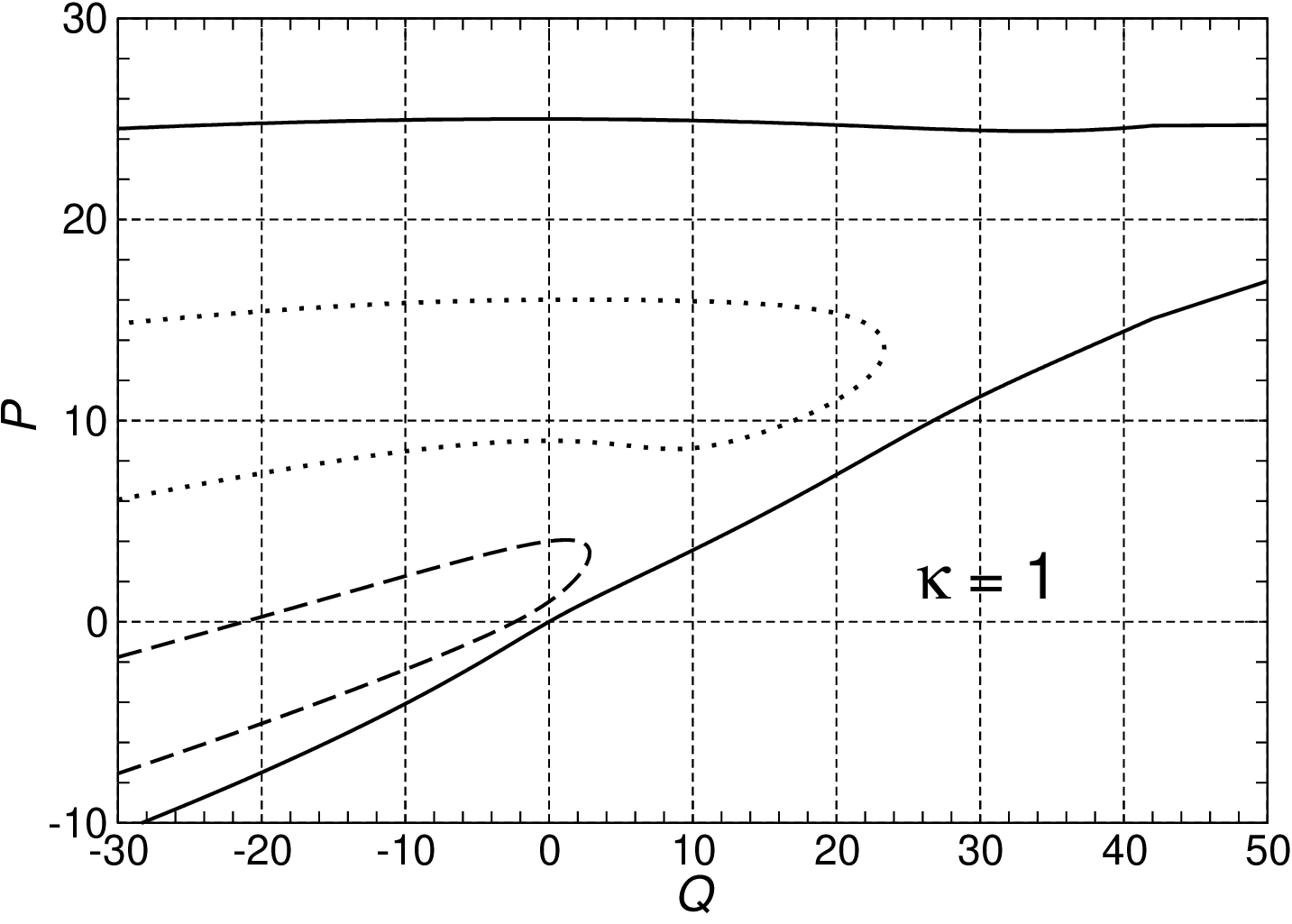}
\caption{The same as in Fig.~1 at $\,\kappa=1$. The main difference is other 
combinations of the coalesced modes. 
In particular, the modes $\,n=0$ and $\,n=5$ join smoothly at 
$\,{\cal Q}\simeq 70$.} 
 \end{figure}
\vspace{200mm}
\begin{figure}[t]
 \includegraphics[width=80mm]{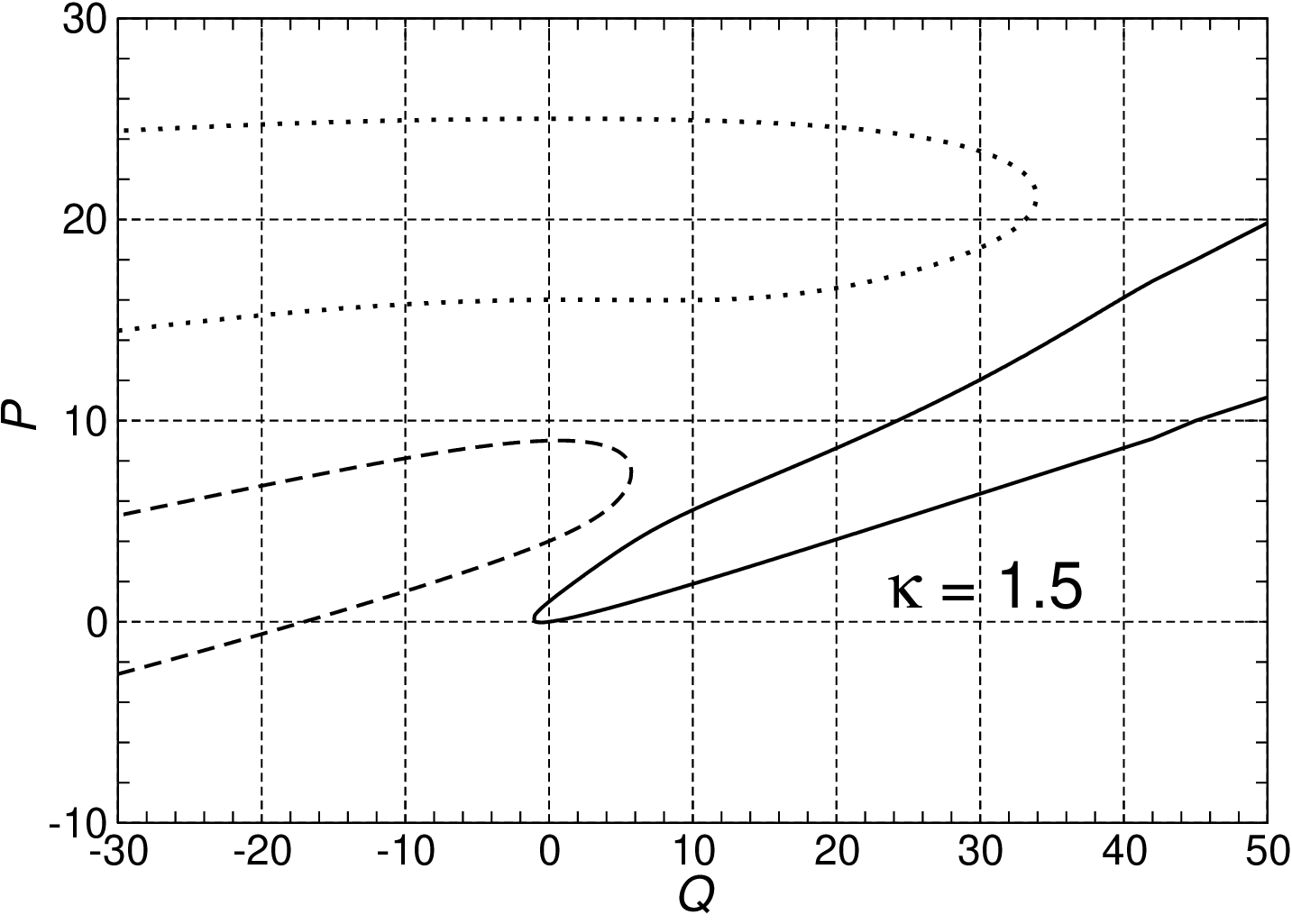}
 \caption{The same as Fig.~1 at $\,\kappa=1.5$. 
The main difference is that real parts of the modes $\,n=0$ and $\,n=1$ 
are coalecsed at $\,{\cal Q}< -1$, and diverge at 
$\,{\cal Q}\rightarrow\infty$.} 
 \end{figure}

\newpage
\section{THE BUNCH EIGENTUNES}

Obtained functions $\,{\cal P}_n(\cal{Q)}\,$ have to be imaged into the plane 
$\,(q_0,\hat\nu)\,$ with help of the transformations 
\begin{eqnarray}
 \hat\nu_{\pm n} = \frac{\Delta Q}{2}\pm\sqrt{\frac{\Delta Q^2}{4}
+{\cal P}_nQ_0^2}, 
 \qquad q_0 = \frac{Q_0^2{\cal Q}}{\hat\nu_{\pm n}}
\end{eqnarray}
which follow from Eq.~(7).
Any point of the family is the real eigentune of the bunch with given SC tune 
shift.
They form several lines representing tunes of different head-tail modes which 
depend on the wake frequency. 

As it has been shown above, $\,{\cal P}_n=n^2$ at $\,{\cal Q}=0$. 
Corresponding solutions are \cite{Bu1}:
\begin{eqnarray}
 \hat\nu_{\pm n} = \frac{\Delta Q}{2}\pm\sqrt{\frac{\Delta Q^2}{4}+n^2 Q_0^2}, 
\qquad {\rm at} \qquad  q_0=0 
\end{eqnarray}
At $\,\Delta Q=0$, it results in $\,\hat\nu_m=mQ_{s0}$ where $\,m=\pm n$.
Such oscillations are known as multipoles.
Any eigenmode of the bunch is a combination of several multipoles.
Nevertheless, the multipole numbers $\,m$ can be used for the classification 
of the lines in the plots. 
For example, a line index $\,\,m=-2$ means that the line passes through the 
point $\,\hat\nu=-2\,\Omega_{s0}$ at $\,q_0=0$ and $\,\Delta Q=0$.
An important role in this consideration play coalescing curves. 
We will use the symbol $\,m=(m_1,m_2)$ to mark the coalescence of the lines of 
kinds $\,m_1$ and $\,m_2$.

Using Figs.~1-3 and Eq. (11), one can get the bunch tunes at selected wake 
frequency and different $\,\Delta Q$.
Some results are plotted in Figs.4-6 where the solid or the dashed lines 
are used repeating appearance of the origin blocks of Figs.~1-3. 
Different colors are used in the resulting figures for different SC tune 
shifts: $\,\Delta Q = 0$ (blue), or $\,\Delta Q/Q_{s0}=1.5$ (green), 
or $\,\Delta Q/Q_{s0} = 3$ (red).
The extreme points of the lines characterized by the relation $\,dq_0=0$ 
highlight beginning of the instability regions. 

The case $\,\kappa=0.5$ is represented in Fig. 4.
It bears the great resemblance to the case of $\,\kappa=0$ which has been 
considered in Ref. \cite{Ba3} with help of the square wake model.
In particular, the blue lines form the chart which is well known 
as the TMCI tunes without space charge $\,\Delta Q=0$ (see e.g. \cite{Ng1}).
At $\,\kappa=0.5$, the instability appears because of coalescence of 
the modes $\,m=0$ and $\,m=\pm 1$ having the threshold $|q_0|/Q_{s0}$ = 0.81 
(0.57 at $\,\kappa=0$ or with the square wake model).

 \begin{figure}[t]
 \includegraphics[width=80mm]{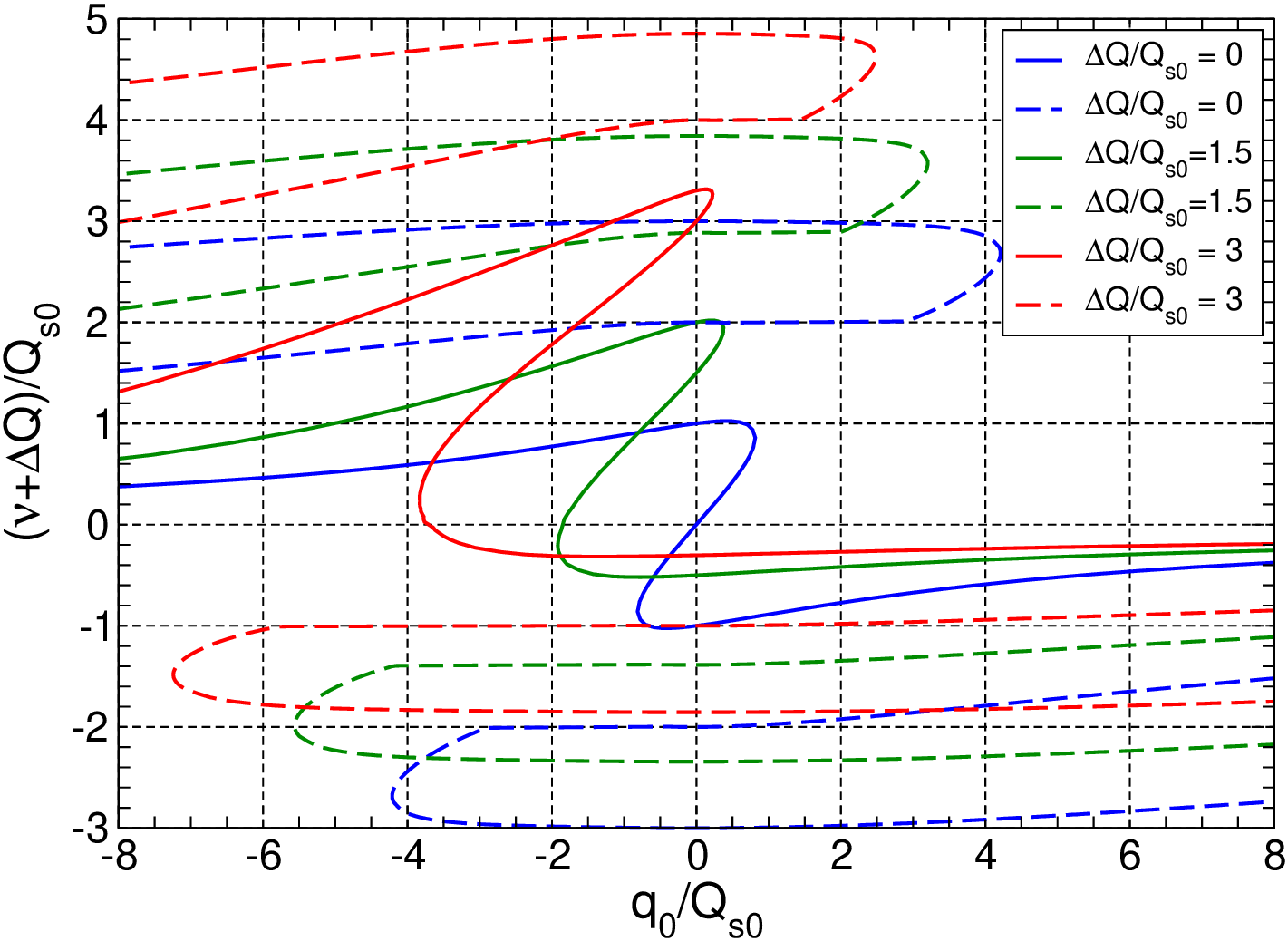} 
\caption{The bunch eigentunes with different $\,\Delta Q$ at $\,\kappa=0.5$  } 
 \end{figure}

\begin{figure}
 \includegraphics[width=80mm]{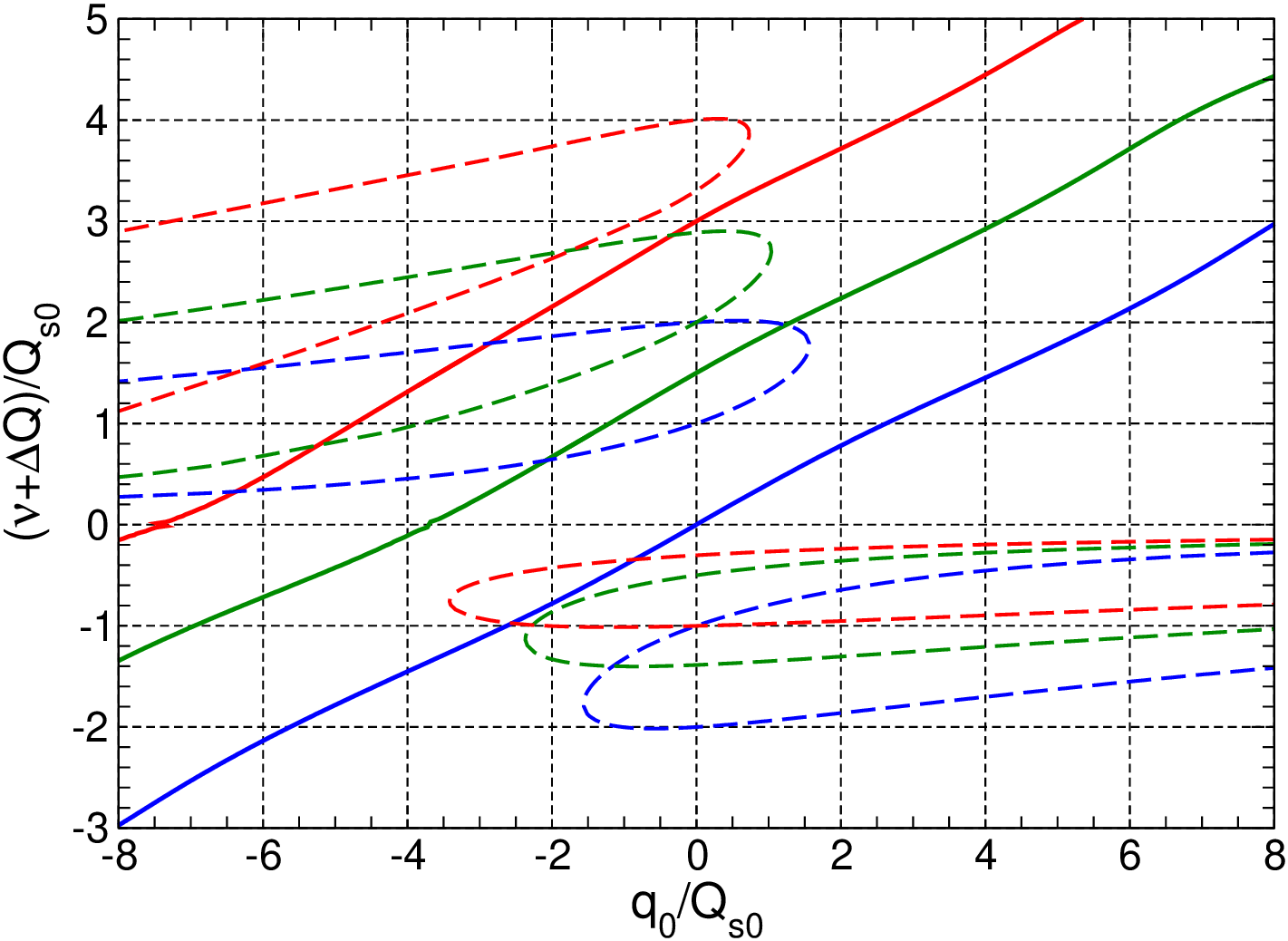}
\caption{The bunch eigentunes with different $\,\Delta Q$ at $\,\kappa=1$.
The same legends as in Fig.~4 are used.}
 \end{figure}

\begin{figure}
 \includegraphics[width=80mm]{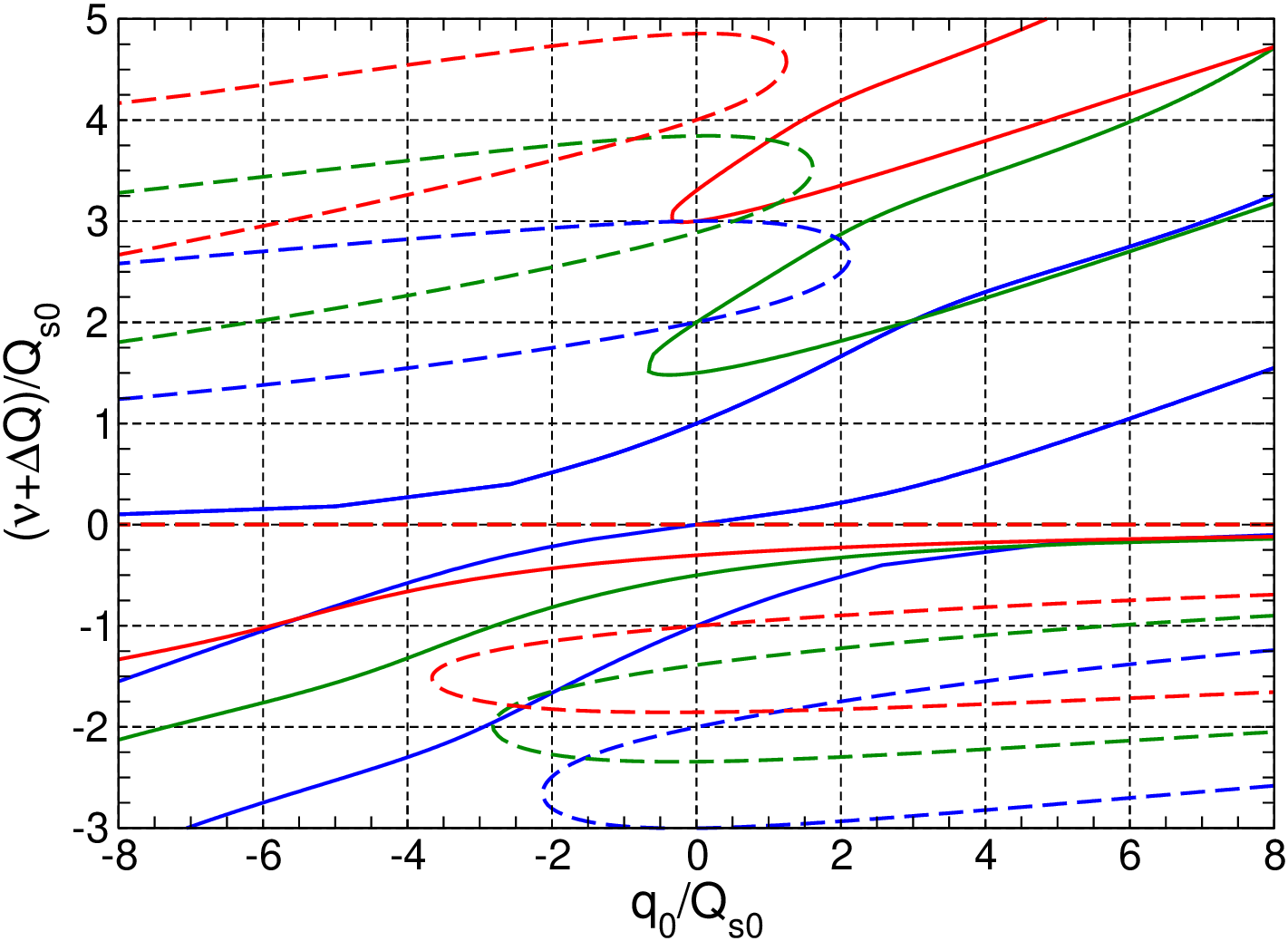}
\caption{The bunch eigentunes with different $\,\Delta Q$ at $\,\kappa=1.5$ 
The same legends as in Fig.~4 are used.  } 
 \end{figure}

All the lines move to the left-up at increasing $\,\Delta Q$.
It means formally that the TMCI threshold increases in modulus tending to
$-\infty$ at $\,q_0<0$, and decreases tending to 0 at $\,q_0>0$. 
The coinciding results have been obtained with the square wake model 
\cite{Ba3}.
However, only the case $\,q_0<0$ has a physical sense for the oscillating wake 
given by Eq. (5).
It means that the SC tune shift stabilizes the TMCI, at least 
at $\,\kappa \le 0.5$.
It is necessary to take into account that the mode $\,m=(-2,-3)$ depends less 
on $\,\Delta Q$ and moves slower to the left than the mode $\,m=(0,-1)$.
Therefore it is the most unstable at $\,\Delta Q/Q_{s0}<\sim 5-6$.
It has been stated in Ref.~\cite{Ba3} at $\,\kappa=0$, and remains in force at 
the moderate value of $\,\kappa$.   

Similar picture takes place at $\,\kappa=1$ (Fig.~5).
The main difference is that the instability appears because of a coalescence
of the modes 
$\,m=-1$ and $\,m=-2$ which effect has the threshold $\,q_0/Q_{s0}=-1.56$ at 
$\,\Delta Q=0$ (the dashed blue lines).
The mode $\,m=0$ could participate only by the coalescence with the mode 
$\,m=-5$ which is possible at much higher amplitude of the wake beyond the 
graph.

Very different picture has a place at $\,\kappa=1.5$ as it follows from Fig.~6.
At $\,\Delta Q=0$, the instability arises by a coalescence of the modes 
$\,m=-2$ and $\,m=-3$ having the threshold $\,q_0/Q_{s0}=-2.13$ 
(dashed blue lines).
However, the lower modes $\,m=0$ and $\,m=1$ come in to play at 
$\,\Delta Q\ne 0$ resulting in the instability with the threshold 
$\,q_0/Q_{s0}=-0.66$ at $\,\Delta Q/Q_{s0}=1.5$ and $\,q_0/Q_{s0}=-0.34$ 
at $\,\Delta Q/Q_{s0}=3$.
The appearance of positive multipoles as well as the lowering of the TMCI
threshold at increasing $\,\Delta Q$ is the characteristic of a positive 
wake \cite{Ba3}.
It is an expected effect in the case because the wake is mostly positive 
in the bunch at $\,q_0<0$ and phase advance $\,1.5\pi$.

Evolution of the spectral lines $\,m=0$ and $\,m=1$ at $\,\kappa=1.5$ 
and increasing $\,\Delta Q$ is represented in more details in Fig.~7.
The blue lines, showing their initial position at $\,\Delta Q=0$,
move toward each other to meet at $\,\Delta Q/Q_{s0}\simeq 0.3$
and to split up in other direction immediately after that.
The magenta lines show their position and shape at $\,\Delta Q/Q_{s0}=0.35$. 
It is seen that very narrow region of instability $\,-3.5<q_0/Q_{s0}<-3$ 
arises in this case.
However, it expands very rapidly at increasing $\,\Delta Q$ obtaining the location 
$\,-7.0<q_0/Q_{s0}<-1.3$ at $\,\Delta Q/Q_{s0}=0.7$ (brown lines), 
and $\,-11.4<q_0/Q_{s0}<-0.7$ at $\,\Delta Q/Q_{s0}=1.5$ (green lines)
Nevertheless, the bunch does take stable in the left ot this area because 
threshold of the mode $\,m=(-1,-2)$ is essentially lower as it follows 
from Fig.~6.

\begin{figure}[t]
 \includegraphics[width=80mm]{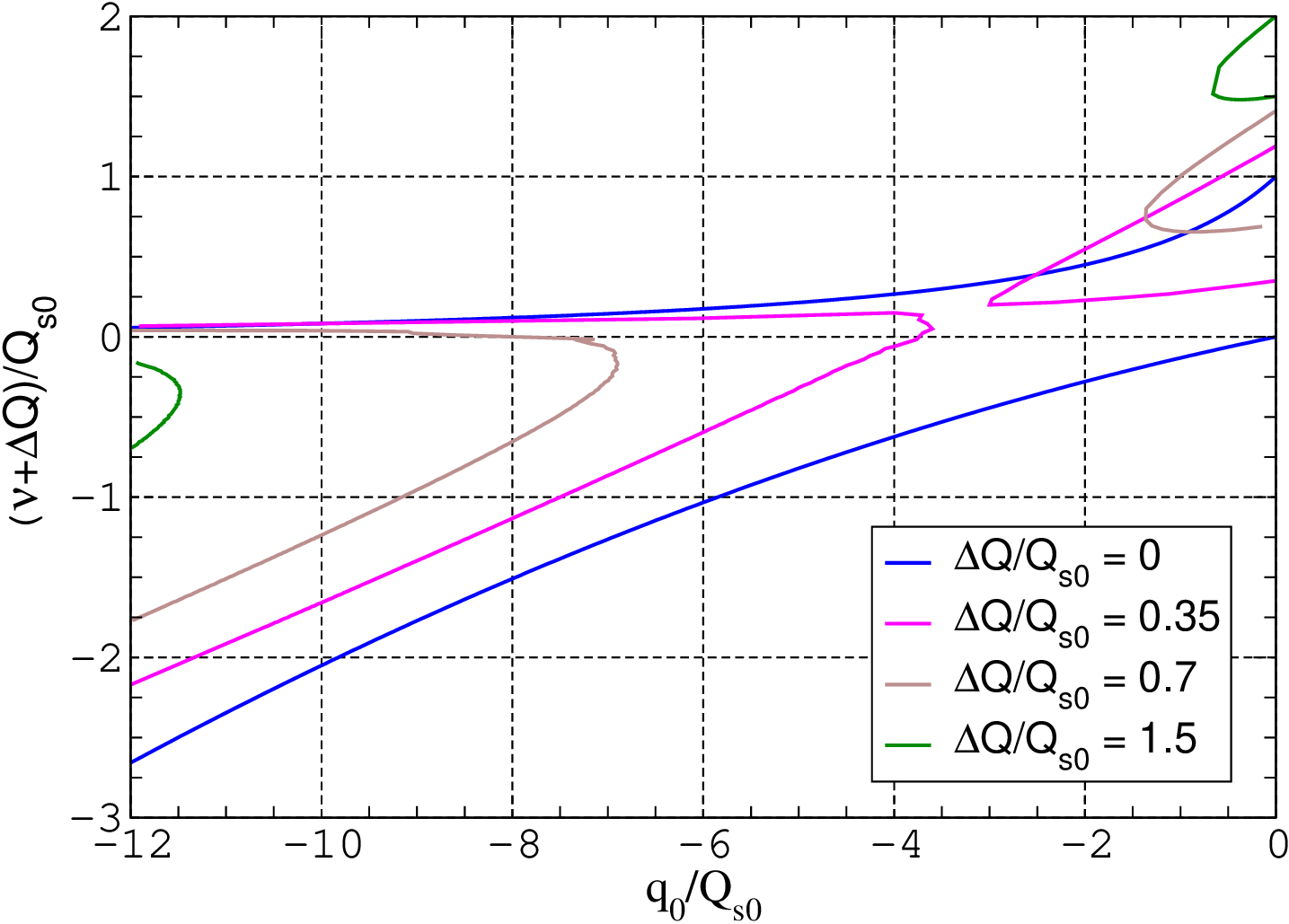}
 \caption{The lowest bunch mode $\,m=(0,1)$ as a function of the 
wake amplitude at $\,\kappa=1.5$ and different $\,\Delta Q/Q_{s0}$. 
At $\,\Delta Q/Q_{s0}>0.35$, the mode becomes unstable at sufficiently large 
$\,q_0$ but obtains the stability again at the further increase of the wake.
However, the bunch can remain unstable because of excitation of the 
mode $\,m=(-2,-3)$.  }
 \end{figure}

\newpage
\section{THE INSTABILITY THRESHOLD}

The developed method allows to find the TMCI threshold with any set of 
parameters.
Some results are represented in this section. 

Dependence of the threshold on the wake frequency is shown in Fig.~8 
at $\,\Delta Q/Q_{s0}=0$, 1.5, and 3.
One can see the oscillations of the threshold and the slow growth of its average 
value with frequency at $\,\kappa>1.5$, which effects have to be expected. 
However, the most important conclusion is that the SC stabilizes the TMCI at 
$\,\Delta Q/Q_{s0}<1$ and destabilizes it at $\,\Delta Q/Q_{s0}>1.5$.
The transition takes place rather rapidly at $\,\Delta Q\simeq 1.25$.

\begin{figure}
 \includegraphics[width=80mm]{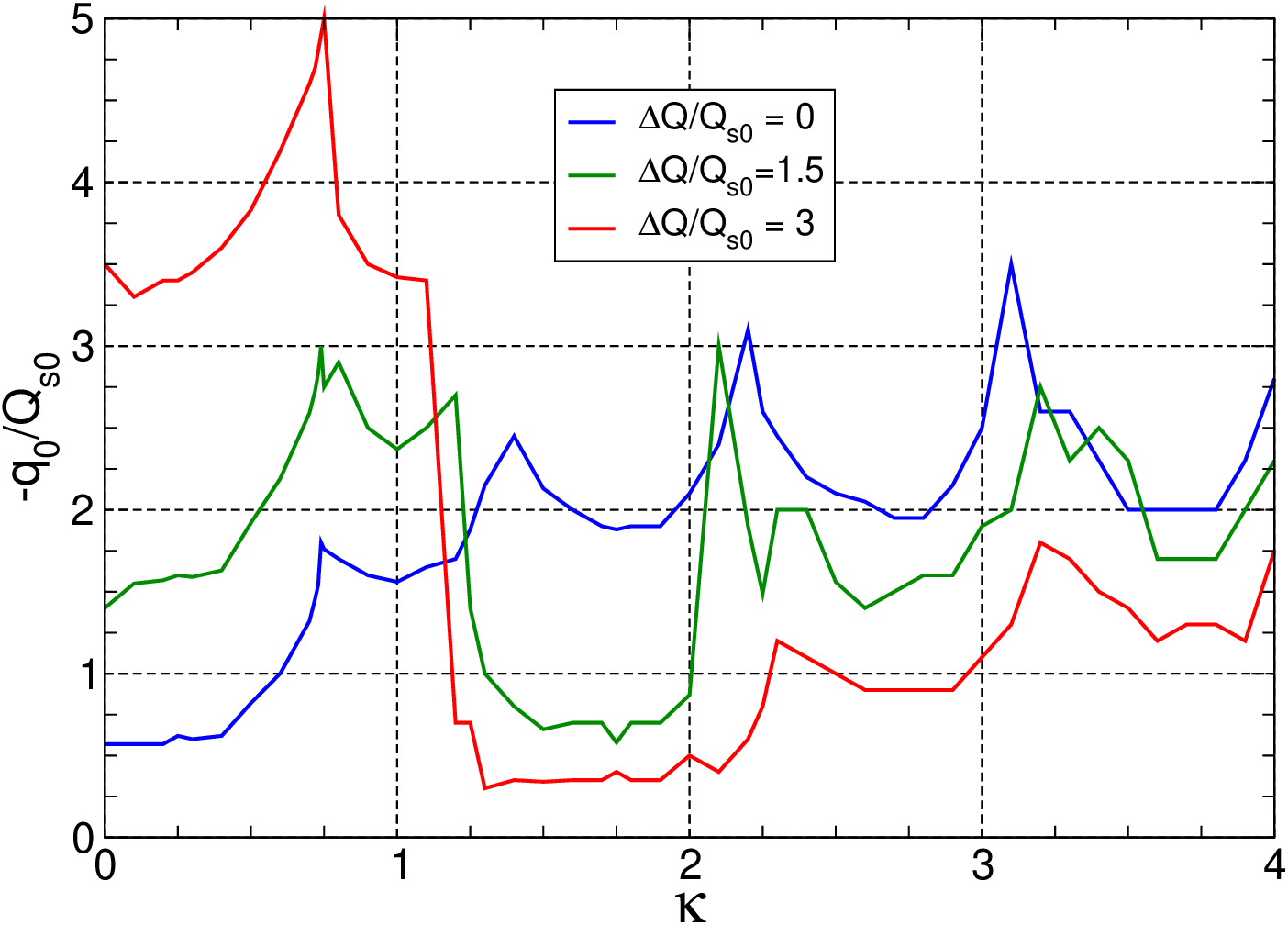}
 \caption{The TMCI threshold against the wake frequency at relatively low
space charge tune shift. The shift stabilizes the bunch at $\,\kappa<1$ and 
destabilizes it at $\,\kappa>1.5$.} 
 \end{figure}
\begin{figure}[t]
 \includegraphics[width=80mm]{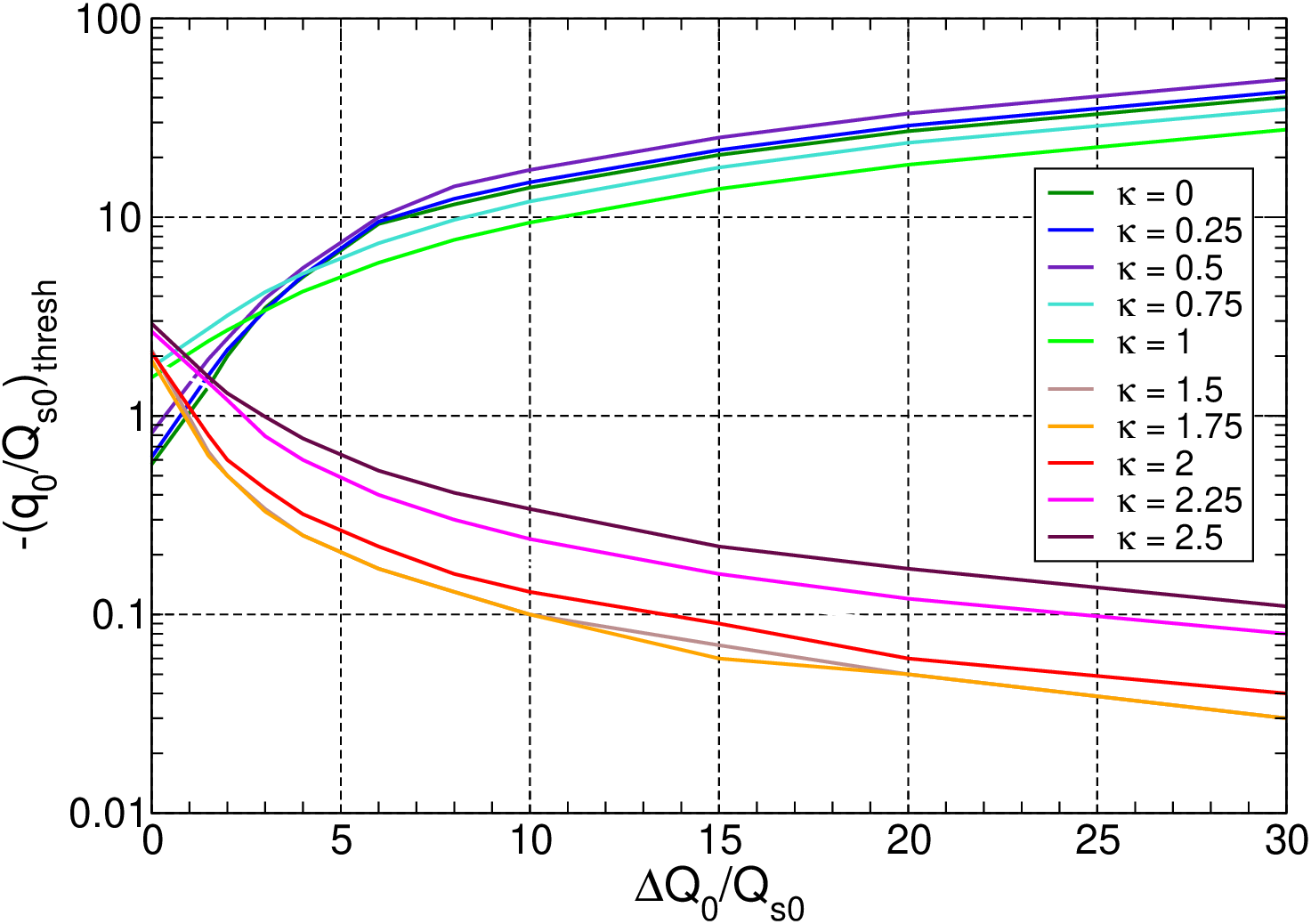}
 \caption{The TMCI threshold against the SC tune shift at different wake 
frequencies.
The parts with frequencies $\,\kappa<1$ and $\kappa>1.5$ are clearly separated.}. 
 \end{figure}
\newpage
Dependence of the threshold on the SC tune shift is shown in Fig.~9 over a 
wider range of the parameters.
The most remarkable effect is that the lines representing different wake 
frequencies are distinctly broken down into 
two groups: with upper threshold at $\,\Delta Q/Q_{s0}<1$ and with lower one at 
$\,\Delta Q/Q_{s0}>1.5$.
They have different asymptotic behavior: $\,q_0\sim\Delta Q$ in the upper group
and $\,q_0\sim Q_{s0}^2/\Delta Q$ in the lower one.
Such a behavior can be obtained with help of Eq.~(11) at the assumption
\begin{equation}
|{\cal P}|\ll (\Delta Q/2Q)^2.
\end{equation}
With sign '--' before the root, it gives
\begin{equation}
\hat\nu\simeq-\frac{{\cal P}Q_{s0}^2}{\Delta Q}, \qquad
q_0\simeq-\frac{\cal Q}{\cal P}\Delta Q
\end{equation}
According these equations, the value of $\,q_0$ varies when the point 
$({\cal Q},{\cal P})$ moves along one of the curves like shown in Fig. 4 
or Fig. 5. 
The TMCI threshold obtained by the last Eq.~(14) appears in the point where 
the condition $\,dq_0 = 0$ is fulfilled that is 
$\,d{\cal P}/{\cal P} = d{\cal Q}/{\cal Q}$. 
It is the point of tangency of the curve with the straight line 
${\cal P} = k_{tg}{\cal Q}$.
Therefore the asymptotic expression for the threshold is:
$$
q_0 = k_{th} \Delta Q
$$
where the constant depends on the wake frequency $\,\kappa$ as it is represented 
in Table I

Another possibility follows from Eq.~(11) and Eq.~(13) with sign '+' before the 
root:
$$
\hat\nu\simeq\Delta Q,\qquad q_0\simeq \frac{{Q_{s0}^2\cal Q}}{\Delta Q} 
$$
According these equations the value of $\,q_0$ varies when the point 
$({\cal Q},{\cal P})$ moves along one of the curves like shown in Fig. 6.
The relation $\,dq_0=0$, that is $\,d{\cal Q}=0$, should be applied 
additionally to find the instability threshold.
It follows from Fig.~6 that the corresponding value of the 
$\,{\cal Q}_{ex}\simeq -1.1$ at $\kappa=1.5$. 
Other values are represented in Table II.
The results are in a conflict with the output of Ref.~\cite{Bu1}
where it is predicated that the space charge has almost no affects 
on the TMCI threshold at $\,\kappa=2$. 

\vspace{5mm}
\begin{table}[h!]
\begin{center}
\caption{Dependence of the asymptotic coefficients on the wake frequency: 
$q_0=k_{tg}(\kappa)\Delta Q$}
\vspace{5mm}
\begin{tabular}{|l|c|c|c|c|c|c|c|}
\hline 
$~~\kappa$~~& ~~0~~  & ~~0.25~~ & ~~0.5~~ & ~~0.75~~ & ~~1~~ \\
\hline
$  ~~k_{tg}~~ $  & -1.34~~&-1.43~~&-1.65~~&-1.17~~&-0.92~~\\
\hline
\end{tabular}
\end{center}
\end{table}
\begin{table}[h!]
\begin{center}
\caption{Dependence of the asymptotic coefficients on the wake frequency: 
$q_0=Q_{s0}^2{\cal Q}_{ex}(\kappa)/\Delta Q$}
\vspace{5mm}
\begin{tabular}{|l|c|c|c|c|c|c|c|}
\hline 
$~~\kappa$~~& ~~1.5~~  & ~~1.75~~ & ~~2~~ & ~~2.25~~ & ~~2.5~~ \\
\hline
$  ~~{\cal Q}_{ex}~~ $  & -1.1~~&-1.1~~&-1.3~~&-2.4~~&-3.4~~\\
\hline
\end{tabular}
\end{center}
\end{table}
The data submitted in Fig. 9 at $\,\kappa\le 1$ are in a good agreement with results 
	of the paper \cite{Ba2} where synchrotron oscillations in a parabolic potential well 
	have been studied using the standard expansion technique.   
	TMCI in the parabolic well has been considered also in recent paper \cite{Bu2} on the 
	base of the asymptotic equations at $\,\Delta Q\gg Q_{s0}$.
	The coincidence of the plots like shown in Fig. 8 confirms that the TMCI threshold is 
	not very sensitive to used models of the bunch and of the potential well.
	The thresholds presented in Fig. 9 at $\,\kappa>1.5$ are in a satisfactory quantitative
 	compliance with the "non-vanishing TMCI" of Ref. \cite{Bu2} where the square well model 
	has been used also but other algorithm of the calculations has been applied.
  
\section{TRANSITIONAL WAKE FREQUENCY.}
The broad blank area between the top and bottom parts of Fig. 9 is assigned 
for the wake of the frequency $\,1<\kappa<1.5$.
However, the transition is not a smooth process in the case.
Actually, at constant SC tune shift and increasing wake, 
the instability can arise, disappear, appear again in other form, etc.  
This statement is illustrated below on the example of the wake with 
$\,\kappa=1.25$.

\begin{figure}[t]
 \includegraphics[width=80mm]{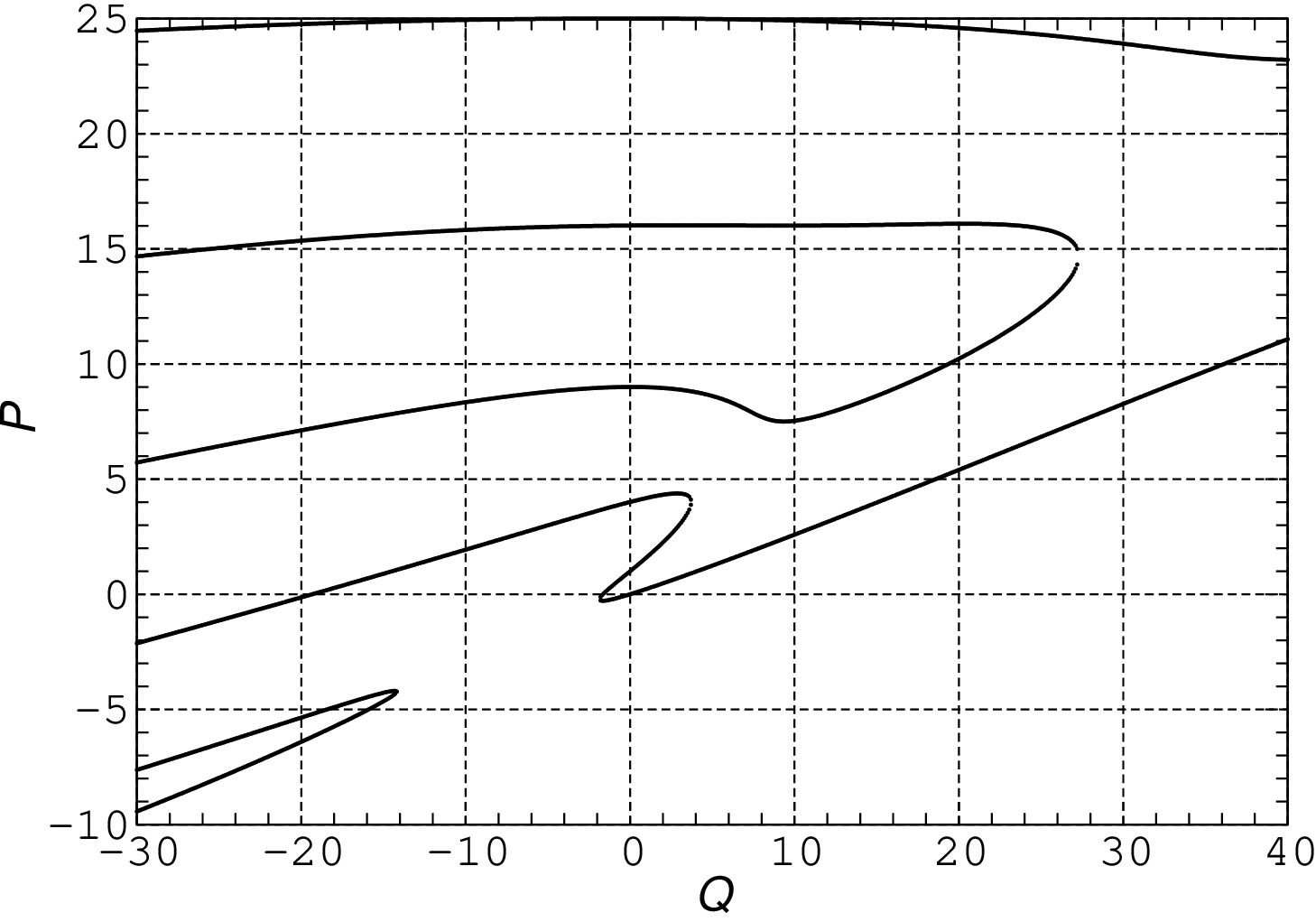}
\caption{Several lowest real eigennumbers of Eq.~(6) and Eq.~(8) at 
$\,\alpha=0,\;\kappa=1.25$.
In comparison with Figs.~2 and 3, there is an additional curl of the lower line} 
 \end{figure}

\begin{figure}
\vspace{10mm}
 \includegraphics[width=80mm]{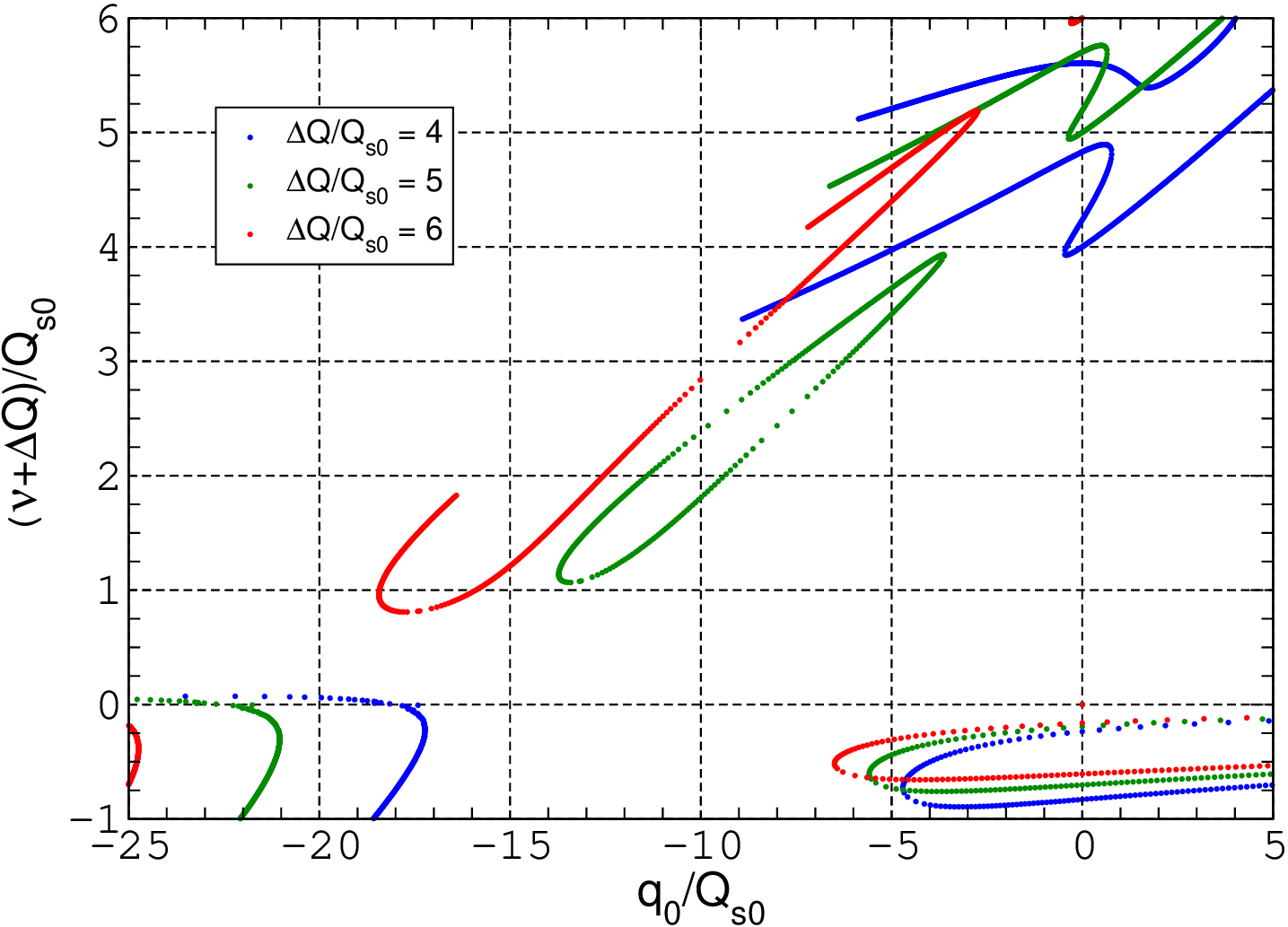}
\caption{The bunch eigentunes at the wake parameters $\,\alpha=0,\,\kappa=1.25$. 
The eigentunes at $\,\Delta Q/Q_{s0}=4$ are shown by dotted blue lines the lower of
them representing the mode $\,m=(-2,-3)$.
There are the islands of stability at $\,\Delta Q/Q_{s0}=5$ and 6 (green and red ovals).} 
\end{figure}

Several lowest eigennumbers of Eq.~(8) are shown on Fig.~10. 
The distinction from Figs.~1-3 consists in the appearance of the additional 
curl with the branches starting from the point 
$\,{\cal Q}\simeq -14,\,{\cal P}\simeq -4$ and steering to the left and down. 
It leads to appearance of the specific areas on the plane $(q_0,\hat\nu)$ 
which are shown on Fig.~11.
Like Fig.~7, instability of the mode $\,m=(0,-1)$ appears not at once
but at sufficiently  large SC tune shift.
For example, with $\,\Delta Q/Q_{s0} =1$, the instability arises 
at $\,q_0/Q_{s0}\simeq-3$.
However, it disappears already at $\,q_0/Q_{s0}\simeq-3.1$  
that is,  in practice, the instability is behind the scenes at such tune shift.
Other examples are given in Fig.~11 where the larger tune shifts 
are considered.
It is seen that the instability band extends very quickly at increasing 
$\,\Delta Q$ reaching approximately the area $-17<q_0/Q_{s0}<-0.5$ 
at $\,\Delta Q/Q_{s0}=4$
This case is shown by the blue lines in the top-right corner and in the 
bottom-left one of Fig.~11.
It should be noted that  the mode $\,m=(-2,-3)$ has the threshold $\,q_0/Q_{s0}=-4.7$
at $\,\Delta Q/Q_{s0}=4$ which tune is shown by the blue dotted line 
in the bottom-right corner of Fig.~11.
 Therefore, with such SC tune sift, the broad picture is:
\\
\\

The bunch is stable at $\,|q_0/Q_{s0}|<0.5$;

The mode $\,m=(0,-1)$ becomes unstable at 

$0.5<|q_0/Q_{s0}|<4.7$;

Both modes $\,m=(0,-1)$ and $\,m=(-2,-3)$ 

are unstable at $\,4.7<|q_0/Q_{s0}|<17$;

Only the mode $\,m=(-2,-3)$ is unstable at 

$\,|q_0/Q_{s0}|>17$.
\\

At larger SC tune shifts, the picture becomes even more complicated 
because of an appearance of the additional islands of stability of the mode 
$\,m=(0,-1)$.
They are shown in Fig.~11 by green and red ovals related to the cases 
$\,\Delta Q/Q_{s0}=5$ or 6, correspondingly.
As a result, following picture is played e.g. at $\,\Delta Q/Q_{s0}=6$ 
(red lines):
\\
\\

The bunch is stable at $|q_0/Q_{s0}|<0.3$;

The mode $\,m=(0,-1)$ becomes unstable at 

$\,0.3<|q_0/Q_{s0}|<2.8$;

The instability disappears at $2.8<|q_0/Q_{s0}|<6.4$;

The mode $\,m=(-2,-3)$ becomes unstable at 

$6.4<|q_0/Q_{s0}|<18.2$;

Both modes $\,m=(0,-1)$ and $\,m=(-2,-3)$ 

are unstable at $\,18.2<|q_0/Q_{s0}|<24.8$;

Only the mode $\,m=(-2,-3)$ remains unstable at 

$\,|q_0/Q_{s0}|>24.8$;
\\
\begin{figure}[t!]
 \includegraphics[width=82mm]{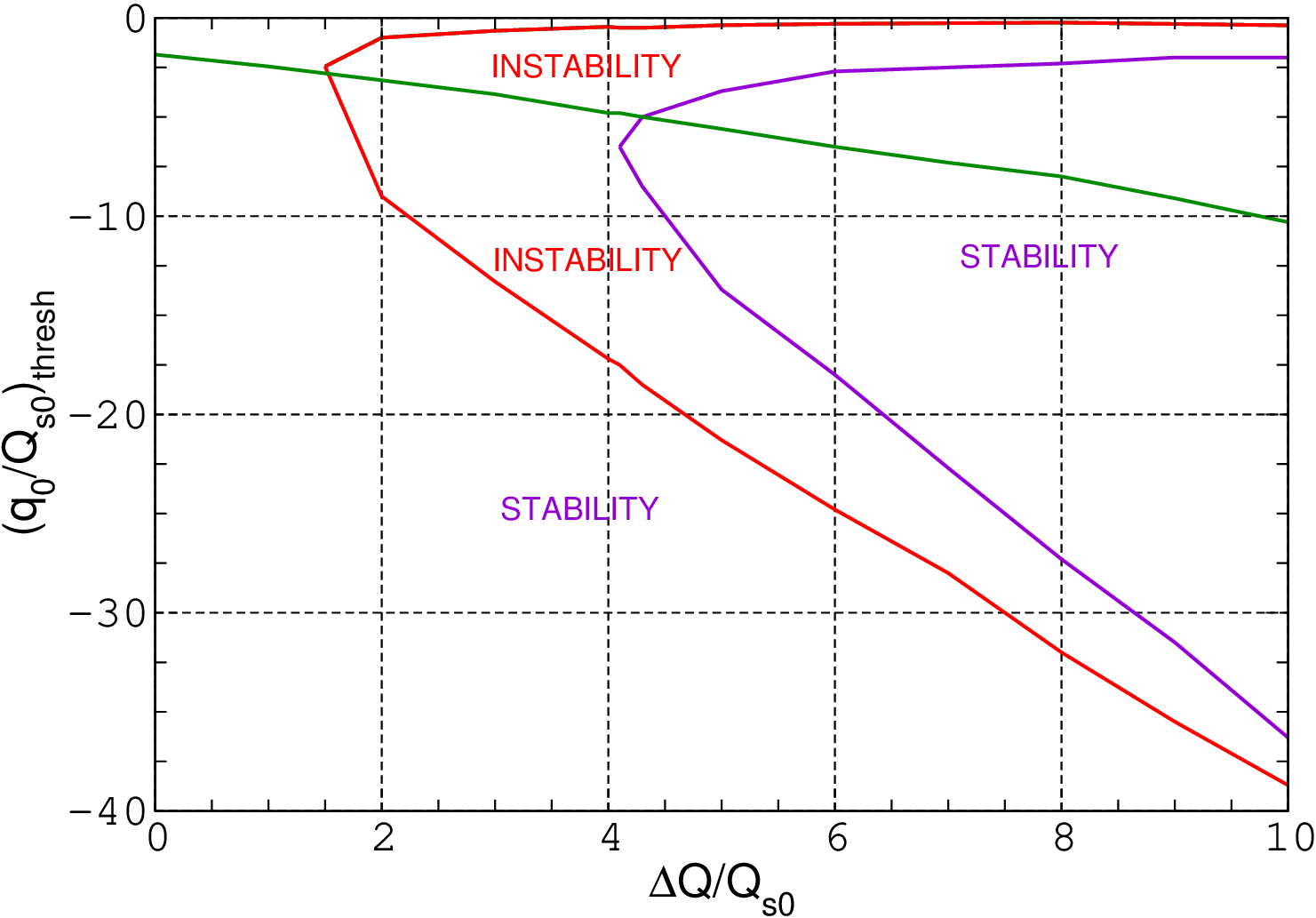}
 \caption{Thresholds of different bunch modes at $\,\kappa=1.25$. 
Red and violet: $\,m=(0,-1)$, green: $\,m=(-2,-3)$.} 
 \end{figure}

The process is illustrated also by Fig.~12.
The mode $\,m=(0,-1)$ may be unstable in the regions between the red lines, 
or between the red and the violet lines but not between the violet lines. 
The legends in the picture concern namely this mode.
Another mode $\,m=(-2,-3)$ is unstable everywhere below the green line.
The asymptotic of this threshold is $\,q_0=-0.9\Delta Q$ 
which expression is quite coordinated with Table II.

Thus, with the oscillating wake, the TMCI threshold can be a nonmonotonic 
function of both the SC tune shift and the WF amlitude.
The result is similar to that found in Ref.~\cite{Bl1} in framework
of the expansion technique. 

\section{Conclusions}

The TMCI threshold of the oscillating wake without space charge 
increases at increase of the wake frequency.  
The space charge effect depends on the wake phase advance on the 
bunch length $\,\phi$.

At $\,\phi<\pi$, the threshold is about proportional to the space
charge tune shift $\,\Delta Q$.

At $\,\phi>1.5\pi$, the threshold is about $\,\propto Q_{s}^2/\Delta Q$
with $\,Q_{s}$ as the synchronron tune.

At $\,\phi\simeq 1.25\pi$, the TMCI threshold can be a non-monotonic 
function of both the space charge tune shift and the wake amplitude.
The variability occurs because different multipole numbers 
are responsible for the instability at different conditions.

The phase advance more of $4\pi$ is not considered in the paper.

\section{Acknowledgment}

Fermi National Accelerator Laboratory is operated by Fermi Research Alliance, 
LLC under Contract No. DEAC02-07CH11395 with the United States Department of 
Energy.


\end{document}